\documentclass[twocolumn,showpacs,amsmath,amssymb,aps,pra]{revtex4}
\usepackage{graphicx}
\usepackage{subfigure}
\usepackage{bm}

\begin{document}

\title{Coincident count rates in absorbing dielectric media}

\author{J. A. Crosse}
\email{jac00@imperial.ac.uk}
\author{Stefan Scheel}
\email{s.scheel@imperial.ac.uk}
\affiliation{Quantum Optics and Laser Science, Blackett Laboratory, 
Imperial College London, Prince Consort Road, London SW7 2AZ, UK}

\date{\today}

\begin{abstract}
A study of the effects of absorption on the nonlinear process of parametric
down conversion is presented. Absorption within the nonlinear medium is
accounted for by employing the framework of macroscopic QED and the Green tensor
quantization of the electromagnetic field. An effective interaction Hamiltonian,
which describes the nonlinear interaction of the electric field and the linear
noise polarization field, is used to derive the quantum state of the light
leaving a nonlinear crystal. The signal and idler modes of this quantum state
are found to be a superpositions of the electric and noise polarization fields.
Using this state, the expression for the coincident count rates for both Type I
and Type II conversion are found. The nonlinear interaction with the
noise polarization field were shown to cause an increase in the rate on the
order of $10^{-12}$ for absorption of $10\%$ per cm. This astonishingly small
effect is found to be negligible compared to the decay caused by linear absorption 
of the propagating modes. From the expressions for the biphoton amplitude it can be 
seen the maximally entangled states can still be produced even in the presence of 
strong absorption.
\end{abstract}


\maketitle
\section{Introduction}
\label{intro}

The nonlinear phenomenon of parametric down conversion (PDC) is ubiquitous
across all areas of quantum optics. The strongly correlated photon pairs
generated by this process have wide applicability in a disparate range of areas;
from fundamental tests of quantum mechanics
\cite{FQMtest1,FQMtest2,FQMtest3,FQMtest4,FQMtest5} to quantum cryptography and
quantum information processing \cite{pdcqi1,pdcqi2,pdcqi3}. As a result, an
extensive body of literature exists on this process (for a necessarily
incomplete selection, see
Refs.~\cite{pdc1,pdc2,pdc3,pdc4,pdc5,pdc6,pdc7,pdc8,pdc9,pdc10}). 
However, despite this continued interest, many features of PDC have not been
studied in detail. As the focus is usually on the interacting electromagnetic
fields, it is common to neglect some of the more complicated aspect of the
nonlinear medium. As a result, absorption is usually ignored.

Absorption, however, plays a key role in all physical (response)
processes as it is an essential requirement of causality. 
Methods for consistently quantizing the electromagnetic field in absorbing
linear electric and magnetic materials have been developed (for a recent review,
see Ref.~\cite{acta}), and recently this formalism has been extended to include
nonlinear processes. In Ref.~\cite{nonlinearH} the authors derive an effective
interaction Hamiltonian to characterize PDC in absorbing media. The result
showed the appearance of extra noise field interactions, with the pump photon
now having the ability to convert not only to electric field modes, but also to
noise polarization modes or a combination of the two. Furthermore, the
Hamiltonian leads to the consistent inclusion of linear absorption as well.
Hence, a full description of the effect of absorption can be obtained using this
formalism.

In this article we will consider the effect of these extra features on the
coincident count rate for Type I and Type II down conversion processes. 
In Sec.~\ref{H}, by studying the evolution of the quantum state of the
signal and idler modes subject to an effective nonlinear interaction
Hamiltonian, we derive the state vector for the outgoing modes.
In Sec.~\ref{Coincident} the state vector will be used to
calculate the coincident count rates for both types of crystal configuration. 
Discussion and some concluding remarks can be found in Secs.~\ref{Dis} and
\ref{Sum}.

\section{The Effective interaction Hamiltonian and the state vector}
\label{H}

We begin by briefly reviewing the concepts of field quantization in absorbing
(linear) dielectric media. The electromagnetic field is expanded in terms of the
classical Green tensor for the Helmholtz operator \cite{acta}. The macroscopic
fields can then be expressed in terms of bosonic operators that describe
collective excitations of the electromagnetic field and the absorbing matter. As
a result the electric field operator becomes
\begin{equation}
\hat{\mathbf{E}}(\mathbf{r}) = \int\limits^{\infty}_0 d\omega\,
\hat{\mathbf{E}}(\mathbf{r},\omega) + \mbox{h.c.}.
\end{equation}
with frequency components
\begin{multline}
\hat{\mathbf{E}}(\mathbf{r},\omega) =
\frac{\omega^2}{c^2\varepsilon_0} \int d^3s\,
\bm{G}(\mathbf{r},\mathbf{s},\omega)\cdot
\hat{\mathbf{P}}_\mathrm{N}(\mathbf{s},\omega)
\\ =i\sqrt{\frac{\hbar}{\varepsilon_0\pi}}\frac{\omega^2}{c^2}
\int d^3s \,\sqrt{\varepsilon''(\mathbf{s},\omega)}
\bm{G}(\mathbf{r},\mathbf{s},\omega)\cdot
\hat{\mathbf{f}}(\mathbf{s},\omega),
\label{E}
\end{multline}
and $\varepsilon''(\mathbf{s},\omega)$ being the imaginary part of the complex
permittivity
$\varepsilon(\mathbf{s},\omega)=\varepsilon'(\mathbf{s},\omega)
+i\varepsilon''(\mathbf{s},\omega)$. The Green tensor (or dyadic Green function)
$\bm{G}(\mathbf{r},\mathbf{s},\omega)$ is the unique solution of the
Helmholtz equation for a point source
\begin{equation}
\bm{\nabla}\times\bm{\nabla}\times\bm{G}(\mathbf{r},\mathbf{s},\omega) -
\frac{\omega^2}{c^2}\varepsilon(\mathbf{r},\omega)
\bm{G}(\mathbf{r},\mathbf{s},\omega)
= \bm{\delta}(\mathbf{r}-\mathbf{s}).
\end{equation}
The frequency components of the noise polarization field
\begin{equation}
\hat{\mathbf{P}}_\mathrm{N}(\mathbf{r},\omega) =
i\sqrt{\frac{\hbar\varepsilon_0}{\pi}\varepsilon''(\mathbf{r},\omega)}\,
\hat{\mathbf{f}}(\mathbf{r},\omega)
\label{P}
\end{equation}
drive the electromagnetic field and are due to absorption processes inside the
dielectric medium. The bosonic operators $\hat{\mathbf{f}}(\mathbf{s},\omega)$
and $\hat{\mathbf{f}}^{\dagger}(\mathbf{s},\omega)$ obey the commutation
relation
\begin{equation}
\left[\hat{\mathbf{f}}(\mathbf{r},\omega),
\hat{\mathbf{f}}^{\dagger}(\mathbf{s},\omega')\right] =
\bm{\delta}(\mathbf{r}-\mathbf{s})\delta(\omega-\omega'). 
\label{comm}
\end{equation}
Finally, the bilinear Hamiltonian  
\begin{equation}
\hat{H}_\mathrm{F} = \int d^3r\int\limits_0^\infty d\omega\,
\hbar\omega\, \hat{\mathbf{f}}^{\dagger}(\mathbf{r},\omega)\cdot
\hat{\mathbf{f}}(\mathbf{r},\omega),
\end{equation}
generates the time-dependent Maxwell equations from Heisenberg's
equations of motion for the electromagnetic field operators. 

\begin{figure}[ht]
\centering
\includegraphics[width=1.0\linewidth]{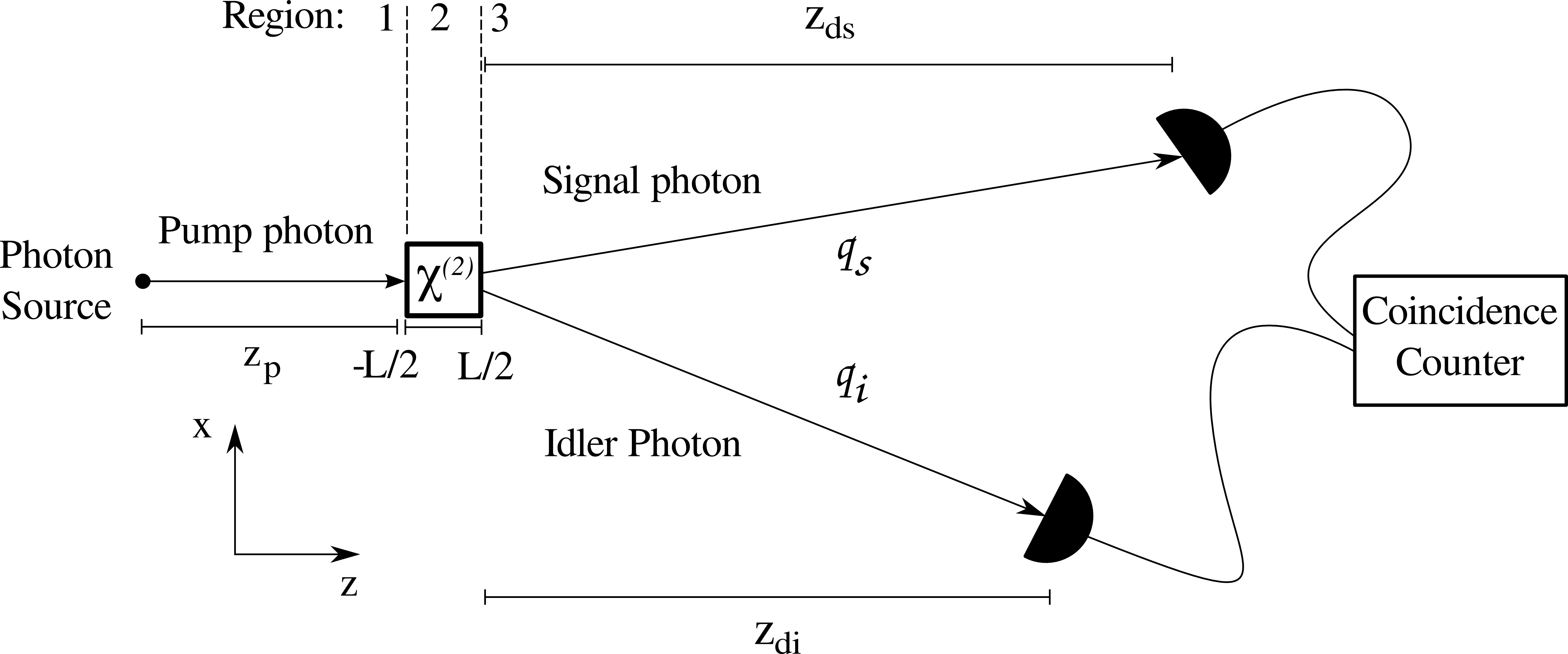}
\caption[Diagram of the setup of a coincident count rate experiment.]{A
$\chi^{(2)}$ crystal of length $L$ is pumped from the left by a continuous,
single-frequency classical plane wave. The resulting signal and idler photons,
with wave vectors $\mathbf{q}_{s}$ and $\mathbf{q}_{i}$, propagate out of the
opposing face of the crystal and are detected in coincidence by detectors at
$\mathbf{r}_{ds}$ and $\mathbf{r}_{di}$.}
\label{Setup} 
\end{figure}

Let us now consider the setup schematically depicted in Fig.~\ref{Setup}.
A $\chi^{(2)}$ crystal of length $L$ (region $2$), surrounded by free space
(regions $1$ and $3$), is pumped from the left by a continuous, single-frequency
classical plane wave with frequency $\omega_{p}$. The classical pump field
approximation asserts that the pump field is of sufficiently high intensity such that it is
negligibly depleted by the nonlinear interaction. Outside the crystal the pump
has wave vector magnitude $q_{p} = \omega_{p}/c$. Inside the crystal the pump
has wave vector magnitude $k_{p} = n(\mathbf{r}_{A},\omega_{p})\omega_{p}/c$,
where $n(\mathbf{r}_{A},\omega)$ is the refractive index of the medium. Owing to
absorption, the refractive index, and hence the wave vector inside the crystal,
is described by a complex number. The coordinate system is taken such that the
pump beam propagates along the $z$-axis. The pump beam enters the crystal at the
planar boundary between regions 1 and 2 located at $z=-L/2$. The interface plane wave
(amplitude) reflection and transmission coefficients are $r_{TEM}^{12}$ and
$t_{TEM}^{12}$, respectively. The pump beam interacts with the medium to produce
signal and idler modes with frequencies $\omega_{s,i}$. The magnitudes of the
wave vectors of these modes inside and outside the crystal are $k_{s,i} =
n(\mathbf{r}_{A},\omega_{s,i})\omega_{s,i}/c$ and $q_{s,i} = \omega_{s,i}/c$,
respectively. The signal and idler modes propagate out of the crystal at the
planar boundary between regions 2 and 3 located at $z=L/2$. The (amplitude) 
reflection and transmission coefficients for this interface are $r_{TE/TM}^{23}$ and
$t_{TE/TM}^{23}$ (note that the outgoing modes are no longer plane waves). For
simplicity the crystal is assumed to be infinitely extended in the $x-y$ plane.
The outgoing modes are detected by coincidence counters located at
$\mathbf{r}_{ds}$ and $\mathbf{r}_{di}$. 

In Ref.~\cite{nonlinearH} we have shown that the effective Hamiltonian that
describes PDC in absorbing media can be written as
\begin{align}
&\hat{H}_\mathrm{int} = \varepsilon_0
\int_{V} d^3r_A\,
\chi_{\alpha\beta\gamma}^{(2)}(\omega_{s},\omega_{i};\omega_{p})
\nonumber\\ 
&\times
\left[ \hat{E}_{s, \alpha}^{\dagger}(\mathbf{r}_{A},\omega_{s},t) +
\mathcal{L}[\varepsilon^\ast(\mathbf{r}_{A},\omega_{s})] 
\hat{P}^{\dagger}_{\mathrm{N}\,s,\alpha}(\mathbf{r}_{A},\omega_{s},t)\right]
\nonumber\\ 
&\times
\left[ \hat{E}^{\dagger}_{i, \beta}(\mathbf{r}_{A},\omega_{i},t)
+\mathcal{L}[\varepsilon^{\ast}(\mathbf{r}_{A},\omega_{i})]
\hat{P}^{\dagger}_{\mathrm{N}\,i,\beta}(\mathbf{r}_{A},\omega_{i},t)
\right]
\nonumber\\ 
&\times
\left[ \hat{E}_{p,\gamma}(\mathbf{r}_{A},\omega_{p},t)
+\mathcal{L}[\varepsilon(\mathbf{r}_{A},\omega_{p})]
\hat{P}_{\mathrm{N}\,p,\gamma}(\mathbf{r}_{A},\omega_{p},t)
\right] + \mbox{h.c.}
\label{Hint}
\end{align}
The greek vector indices, which refer to the polarization of the mode, run
over the three Cartesian coordinates. Over these indices summation convention is
implied. The function $\mathcal{L}[\varepsilon(\mathbf{r}_{A},\omega)]$, given
by
\begin{eqnarray}
\mathcal{L}[\varepsilon(\mathbf{r}_{A},\omega)] = \frac{2}{9\varepsilon_0}
\frac{\varepsilon(\mathbf{r}_{A},\omega)-1}{\varepsilon(\mathbf{r}_{A},\omega)},
\end{eqnarray}
is the local-field correction factor for the noise polarization field, and
$\varepsilon(\mathbf{r}_{A},\omega) = n(\mathbf{r}_{A},\omega)^{2}$ is the
complex linear permittivity within the crystal. The spatial integral is taken
over the volume of the crystal.

The quantum state of light at the output face of the crystal can be found by
evaluating the time evolution of the state vector subject to the interaction
Hamiltonian in Eq.~\eqref{Hint}. Formally, to first order in perturbation
theory, the long-time limit, steady-state quantum state vector is given by
\begin{equation}
|\psi\rangle = |\psi(0)\rangle + \frac{1}{i\hbar}\int_{-\infty}^{\infty} dt\,
\hat{H}(t)|\psi(0)\rangle + \mathcal{O}\left(\hat{H}_{I}(t)^2\right).
\label{wavefn}
\end{equation}
The first order truncation of the series implies that probability of
interaction is sufficiently low as to make the likelihood of a two 
consecutive down conversion processes negligibly small. 

The classical plane-wave pump field can be found from the Green tensor
description of a sheet current source at infinity and is given by
\begin{align}
\hat{E}_{p,\gamma}&(\mathbf{r}_{A},\omega_{p},t) =
\hat{\underline{e}}_{p,\gamma}E_{p}e^{i(q_{p}z_{p}-\omega_{p}t)}t^{12}_{TEM}
(\omega_{p})e^{ik_{p}\frac{L}{2}}\nonumber\\
&\times \big[e^{ik_{p}z_{A}}+e^{-ik_{p}(z_{A}-L)}
r^{23}_{TEM}(\omega_{p})\big]M_{TEM}(\omega_{p}),
\label{p}
\end{align}
written in a form familiar from classical wave optics.
Here, $\hat{\underline{e}}_{p,\gamma}$ is the pump polarization vector and
$z_{p}$ is the distance from the pump source to the input face of the crystal.
The Fresnel coefficients for reflection and transmission of plane waves
at the boundaries are given by
\begin{gather}
r^{21/23}_{TEM}(\omega) = \frac{n(\omega)-1}{n(\omega)+1},\quad
t^{12}_{TEM}(\omega)  = \frac{2}{n(\omega)+1}.
\end{gather}
Multiple scattering within the crystal is accounted for by
\begin{equation}
M_{TEM}(\omega) = \big[1-r^{21}_{TEM}(\omega)r^{23}_{TEM}(\omega)e^{2ik_{p}L}\big]^{-1}.
\end{equation}
All these coefficients can be found in the far-field expansion of the
generalized Fresnel and multiple scattering coefficients of the dyadic Green
tensor for layered media (see Appendix~\ref{App:Green}). 

\begin{figure}[ht]
\centering
\includegraphics[width=0.6\linewidth]{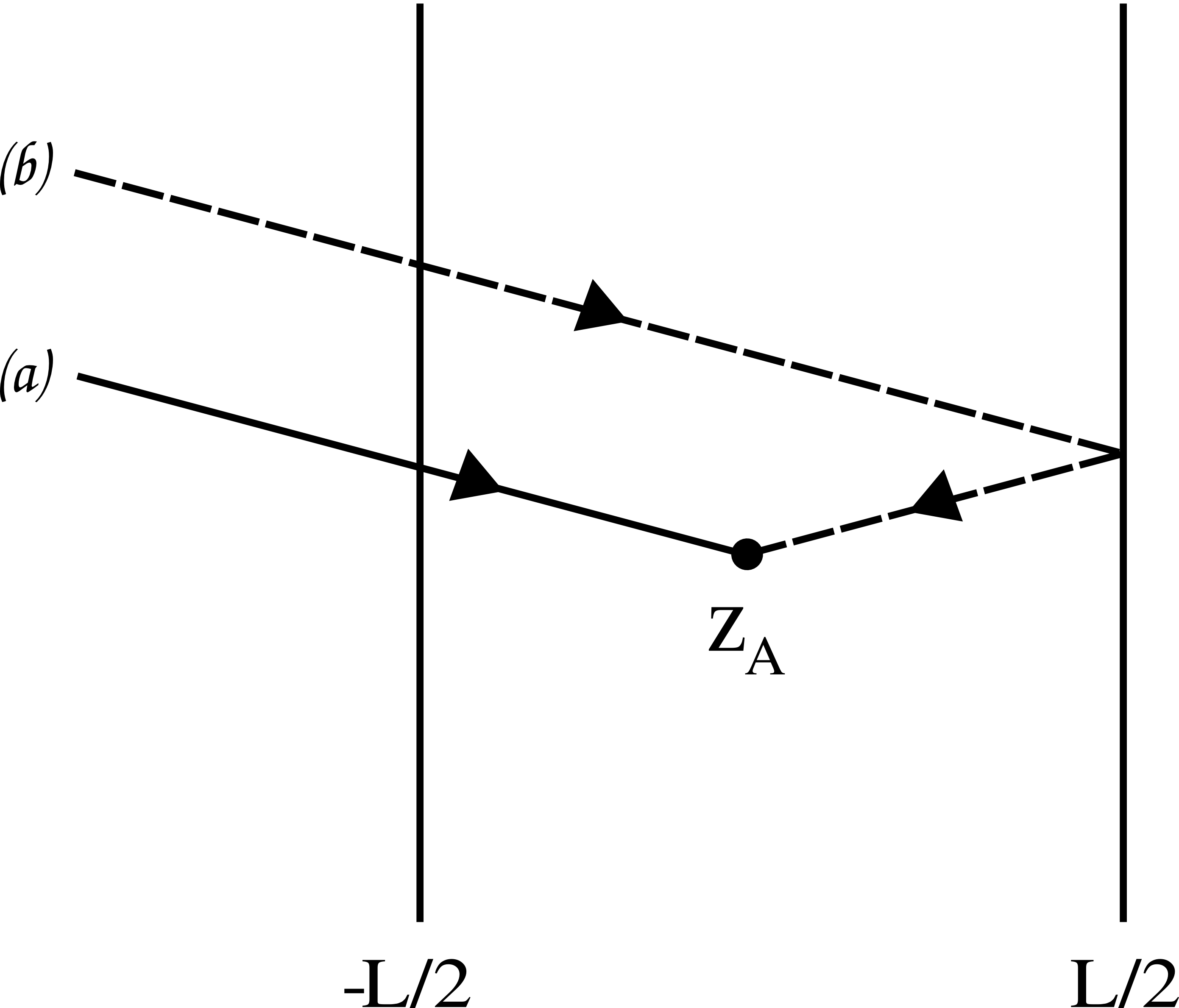}
\caption{The first term in the square brackets of Eq. \eqref{p} corresponds to
propagation path (a), the second term to propagation path (b). Scattering off
two or more interfaces is taken into account by the multiple scattering
coefficient $M_{TEM}(\omega)$.}
\label{Pump} 
\end{figure}

The first term in the square brackets of Eq. \eqref{p} corresponds to photons
that propagate directly to the interaction point. The second term corresponds to
photons that are reflected at the back face of the crystal (see Fig.~\ref{Pump}).
The crystal medium is taken to be in its ground state and hence the
(expectation value of the) noise polarization field at the pump frequency
vanishes. Furthermore, it is assumed that both the electric and noise
polarization fields at the signal and idler frequencies consist of a slowly
varying amplitude and a fast oscillation
\begin{gather}
\hat{E}_{(s,i),\alpha}(\mathbf{r}_{A},\omega_{s,i},t) =
\hat{E}_{(s,i),\alpha}(\mathbf{r}_{A},\omega_{s,i})e^{-i\omega_{s,i}t},
\label{EN}\\
\hat{P}_{\mathrm{N}\,(s,i),\alpha}(\mathbf{r}_{A},\omega_{s,i},t) =
\hat{P}_{\mathrm{N}\,(s,i),\alpha}(\mathbf{r}_{A},\omega_{s,i})
e^{-i\omega_{s,i}t}.
\label{PN}
\end{gather}

After combining Eqs.~\eqref{p}, \eqref{EN}, and \eqref{PN} with the
effective Hamiltonian \eqref{Hint}, the resulting expression is substituted into
Eq.~\eqref{wavefn}. The initial state for both the signal and idler modes are
taken to be the vacuum ($|\psi(0)\rangle = |0\rangle_{s}|0\rangle_{i}$).
Finally, performing the integral over $t$ results in a 
$\delta$-function. This enforces the energy conservation condition
$\omega_{p}=\omega_{s}+\omega_{i}$. One should note that, as the frequency of a
propagating mode is unchanged by a change of medium, this condition is true both
inside and outside the crystal. Thus, one finds the steady-state quantum state
vector at the crystal output as
\begin{align}
&|\psi\rangle  = \mathcal{N}|0\rangle_{s}|0\rangle_{i} +
\frac{4\pi\varepsilon_0E_{p}}{i\hbar}t^{12}_{TEM}(\omega_{p})M_{TEM}(\omega_{p}
)\nonumber\\
&\times \int_{V} d^{3}r_{A}\,d_{\alpha\beta}(\omega_{s},\omega_{i})
\big[e^{ik_{p}z_{A}}+e^{-ik_{p }(z_{A}-L)}r^{23}_{TEM}(\omega_{p})\big]
\nonumber\\
&\times\left[ \hat{E}_{s, \alpha}^{\dagger}(\mathbf{r}_{A},\omega_{s}) +
\mathcal{L}[\varepsilon^\ast(\mathbf{r}_{A},\omega_{s})]
\hat{P}^{\dagger}_{\mathrm{N}\,s,\alpha}(\mathbf{r}_{A},\omega_{s})\right]
\nonumber\\
&\times\left[ \hat{E}^{\dagger}_{i, \beta}(\mathbf{r}_{A},\omega_{i})
+\mathcal{L}[\varepsilon^{\ast}(\mathbf{r}_{A},\omega_{i})]\hat{P}^{\dagger}_{
\mathrm{N}\,i,\beta}(\mathbf{r}_{A},\omega_{i})\right]\nonumber\\
&\times e^{ik_{p}\frac{L}{2}}e^{iq_{p}z_{p}}|0\rangle_{s}|0\rangle_{i},
\label{state}
\end{align}
where $\mathcal{N}$ is a normalization factor that is very close to one. Note
also that a contracted notation for the nonlinear susceptibility has been used:
$2d_{\alpha\beta}(\omega_{s},\omega_{i}) =
\chi_{\alpha\beta\gamma}^{(2)}(\omega_{s},\omega_{i};\omega_{p})\hat{\underline{
e}}_{p,\gamma}$.  

\section{Coincident count rate}
\label{Coincident}

Given a quantum state $|\psi\rangle$, the count rate at the 
detectors located at $\mathbf{r}_{ds}$ and $\mathbf{r}_{di}$ is given by
\cite{pdc2, pdc3}
\begin{multline}
R = \langle\psi
|\hat{E}^{\dagger}_{s,\lambda}(\mathbf{r}_{ds},\omega_{ds})
\hat{E}^{\dagger}_{i,\mu}(\mathbf{r}_{di},\omega_{di})\\
\times\hat{E}_{s,\lambda}(\mathbf{r}_{ds},\omega_{ds})
\hat{E}_{i,\mu}(\mathbf{r}_{di},\omega_{di})|\psi\rangle .
\label{genrate}
\end{multline}
Owing to the number of field operators in $|\psi\rangle$, one can rewrite the
Eq.~\eqref{genrate} as
\begin{equation}
R = A_{\lambda\mu}(A^{\lambda\mu})^{\ast}
\end{equation}
with
\begin{equation}
A_{\lambda\mu}
=\,_{i}\langle 0|_{s}\langle0|\hat{E}_{s,\lambda}(\mathbf{r}_{ds},\omega_{ds})
\hat{E}_{i,\mu}(\mathbf{r}_{di},\omega_{di})|\psi\rangle
\label{amp}
\end{equation}
commonly being referred to as the biphoton amplitude.

Using Eq. \eqref{state} to expand the state vector in Eq. \eqref{amp} gives
\begin{align}
&A_{\lambda\mu} =\frac{4\pi\varepsilon_0E_{p}}{i\hbar}
t^{12}_{TEM}(\omega_{p})M_{TEM}(\omega_{p})e^{iq_{p}z_{p}}e^{ik_{p}\frac{L}{2}}
\nonumber\\
&\times \int_{V} d^{3}r_{A}\,d_{\alpha\beta}(\omega_{s},\omega_{i})
\big[e^{ik_{p}z_{A}}+e^{-ik_{p}(z_{A}-L)}r^{23}_{TEM}(\omega_{p})\big]
\nonumber\\ 
& \times\,_{s}\langle0|\hat{E}_{s,\lambda}(\mathbf{r}_{ds},\omega_{ds})
\nonumber\\
&\times\left[\hat{E}_{s, \alpha}^{\dagger}(\mathbf{r}_{A},\omega_{s}) +
\mathcal{L}[\varepsilon^\ast(\mathbf{r}_{A},\omega_{s})]
\hat{P}^{\dagger}_{\mathrm{N}\,s,\alpha}(\mathbf{r}_{A},\omega_{s})\right]
|0\rangle_{s}\nonumber\\
&\times\,_{i}\langle 0|\hat{E}_{i,\mu}(\mathbf{r}_{di},\omega_{di}) \nonumber\\
&\times\left[ \hat{E}^{\dagger}_{i, \beta}(\mathbf{r}_{A},\omega_{i})
+\mathcal{L}[\varepsilon^{\ast}(\mathbf{r}_{A},\omega_{i})]
\hat{P}^{\dagger}_{\mathrm{N}\,i,\beta}(\mathbf{r}_{A},\omega_{i})\right]
|0\rangle_{i}.
\label{amp2}
\end{align}
The matrix elements can be evaluated using the commutation relations for the
electromagnetic field operators given in Appendix~\ref{App:Comm}. One finds 
that the only contributing terms are those for which $\omega_{ds} = \omega_{s}$ 
and $\omega_{di} =\omega_{i}$. Upon applying the identity 
$\mathrm{Im}\,\bm{G} =(\bm{G}-\bm{G}^{\ast})/2i$, Eq.~\eqref{amp2} becomes
\begin{align}
&A_{\lambda\mu} = \frac{i\hbar
E_{p}}{\pi\varepsilon_0}\frac{\omega_{s}^{2}\omega_{i}^{2}}{c^{4}}
t^{12}_{TEM}(\omega_{p})M_{TEM}(\omega_{p})e^{iq_{p}z_{p}}e^{ik_{p}\frac{L}{2}}
\nonumber\\
&\times \int_{V}
d^{3}r_{A}\,d_{\alpha\beta}(\omega_{s},\omega_{i})
\big[e^{ik_{p}z_{A}}+e^{-ik_{p}(z_{A}-L)}r^{23}_{TEM}(\omega_{p})\big]
\nonumber\\
&\times\bigg[A^{\ast}(\mathbf{r}_{A},\omega_{s})
G_{\lambda\alpha}(\mathbf{r}_{ds},\mathbf{r}_{A},\omega_{s}) -
G^{\ast}_{\lambda\alpha}(\mathbf{r}_{ds},\mathbf{r}_{A},\omega_{s})\bigg]
\nonumber\\
&\times\bigg[A^{\ast}(\mathbf{r}_{A},\omega_{i})G_{\mu\beta}(\mathbf{r}_{di},
\mathbf{r}_{A},\omega_{i}) -
G^{\ast}_{\mu\beta}(\mathbf{r}_{di},\mathbf{r}_{A},\omega_{i})\bigg],
\label{amp3}
\end{align}
with
\begin{equation}
A(\mathbf{r},\omega) = 1-2i\varepsilon_{0}
\varepsilon''(\mathbf{r},\omega)\mathcal{L}[\varepsilon(\mathbf{r},\omega)].
\label{A}
\end{equation}

\begin{figure}[t]
\centering
\includegraphics[width=0.7\linewidth]{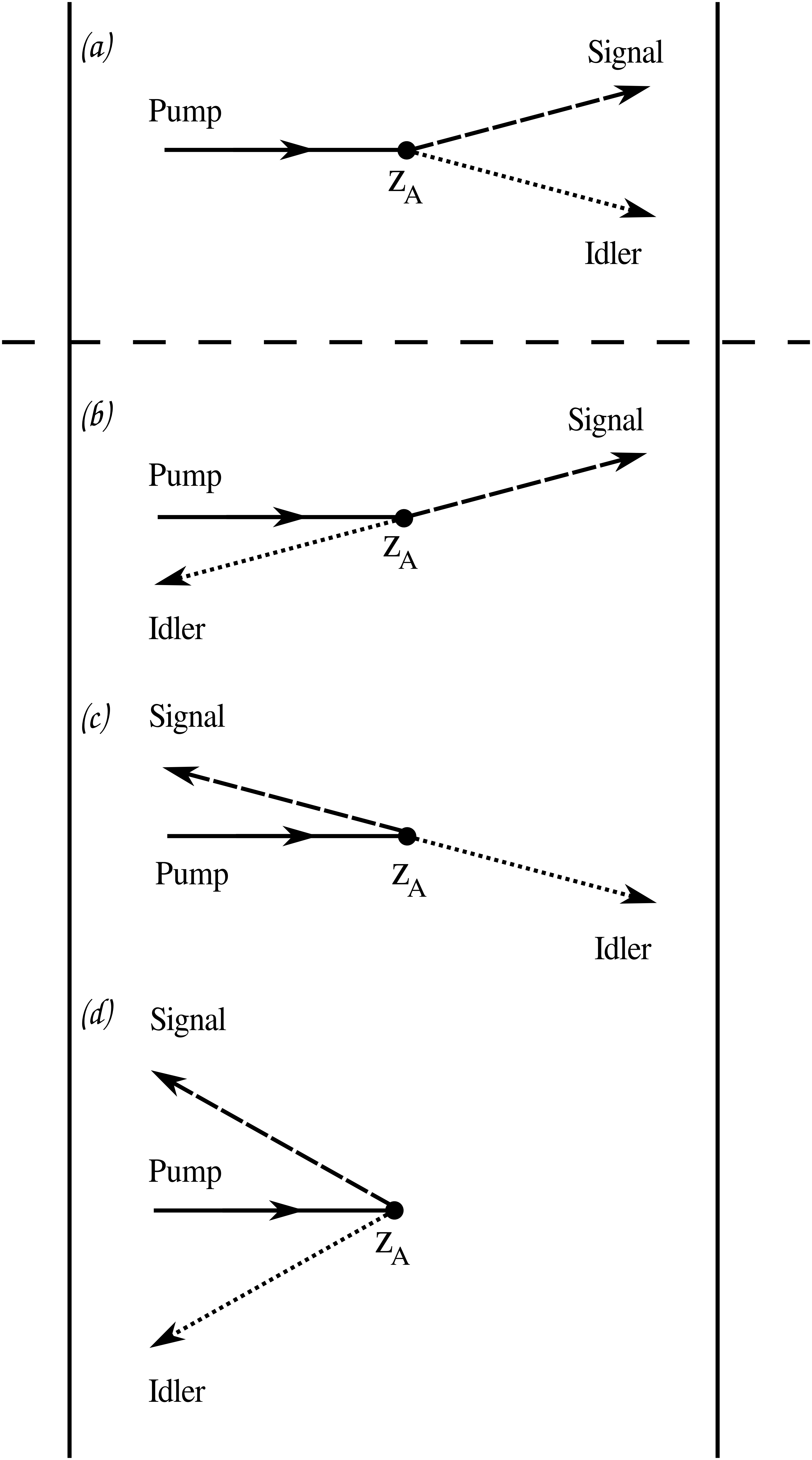}
\caption{The interaction shown in (a) is the forward-scattering term given by
the dominant term $\propto G_{\lambda\alpha}G_{\mu\beta}$. Interactions (b),(c)
and (d) are described by the terms $G^{\ast}_{\lambda\alpha}G_{\mu\beta}$,
$G_{\lambda\alpha}G^{\ast}_{\mu\beta}$ and
$G^{\ast}_{\lambda\alpha}G^{\ast}_{\mu\beta}$, respectively. These involve
back-scattering of one or more of the outgoing modes and can be neglected.}
\label{Dropped} 
\end{figure}

Let us consider the $z$ components of the wave vectors of the three modes [c.f.
Eq.~\eqref{F} in Appendix~\ref{App:Green}]. One can see that the four terms in
Eq.~\eqref{amp3} refer to four different types of directional scattering of the
signal and idler modes. The term proportional to $G_{\lambda\alpha}G_{\mu\beta}$
describes forward scattering where both of the outgoing modes leave in the same
direction as the pump beam. The cross terms and conjugate term correspond to
interactions where back-scattering of one or both of the outgoing modes occurs
(see Fig.~\ref{Dropped}). 

Closer inspection reveals that
this wave vector dependence is contained in a number of exponential terms of the
form $e^{\pm if(k_{z})z}$. In general, for long bulk crystals, the dominant terms 
will be those for which
$f(k_{z}) \approx 0$, when the crystal is said to be longitudinally phase
matched. Away from this condition the contribution to the total amplitude is
proportional to $\mathrm{sinc}[f(k_{z})]$. Hence, even for relatively
small deviations from perfect phase matching, the contribution to the total
amplitude falls sharply. In order to conserve energy and be close to perfect
phase matching, $k_{p}$, $k_{z,i}$ and $k_{z,s}$ must have the same sign.
Furthermore, since $|k_{p}|>|k_{z,s}|,|k_{z,i}|$, the main contribution comes
from the terms where $f(k_{z}) \approx k_{p}-k_{z,i}-k_{z,s}$. All other terms
lie far from phase matching and hence are negligible.
One can see that the terms that are closest to this phase matching condition are
contained in the term proportional to $G_{\lambda\alpha}G_{\mu\beta}$
in Eq.~\eqref{amp3}. Thus, the cross and conjugate terms
can be neglected. In that way, one extracts the relevant contributions to the 
nonlinear process in terms of Green functions, as a generalization but in the 
same spirit as the free-space analysis using plane waves. 

It is worth noting that, although negligible for long bulk crystals, the cross
and conjugate terms may give comparable contributions in layered media
\cite{perina1, perina2}. However, the Green tensor method is well adjusted for 
the study of such systems as the Green tensor for layered media is well known
and hence these effects can easily be investigated.

Further to this, we shall also assume that the linear permittivity is constant
across the crystal, hence $A_{\alpha}(\mathbf{r},\omega)=A_{\alpha}(\omega)$ in
the range of integration. Thus, one can take these prefactors out of the spatial
integral. As a result, Eq.~\eqref{amp3} becomes
\begin{align}
&A_{\lambda\mu} = \frac{i\hbar E_{p}}{\pi\varepsilon_0}
\frac{\omega_{s}^{2}\omega_{i}^{2}}{c^{4}}t^{12}_{TEM}(\omega_{p})
M_{TEM}(\omega_{p})e^{iq_{p}z_{p}}e^{ik_{p}\frac{L}{2}} \nonumber\\
&\times\int_{V} d^{3}r_{A}\,d_{\alpha\beta}(\omega_{s},\omega_{i})
\big[e^{ik_{p}z_{A}}+e^{-ik_{p}(z_{A}-L)}r^{23}_{TEM}(\omega_{p})\big]
\nonumber\\
&\times A^{\ast}(\omega_{s})A^{\ast}(\omega_{i})
G_{\lambda\alpha}(\mathbf{r}_{ds},\mathbf{r}_{A},\omega_{s})
G_{\mu\beta}(\mathbf{r}_{di},\mathbf{r}_{A},\omega_{i})
\label{amp4}
\end{align}
which serves as the starting point for the special cases considered below.

\subsection{Type I down conversion}
\label{Type I}

The Green tensors appearing in Eq.~\eqref{amp4} can be expanded 
in terms of vector wave functions as
given in Appendix~\ref{App:Green}. As the crystal is infinitely extended in
the ($x,y$)-plane, the integral over the transverse components of $\mathbf{r}_{A}$ 
is proportional to a $\delta$-function. Integrating over it
leaves one transverse $\mathbf{k}_{\perp}$-integral and the spatial integral in the
$z$-direction over the length of the crystal. 

If the crystal is cut for Type I conversion the polarization 
unit vectors of the outgoing modes are parallel. Hence the
nonlinear susceptibility has the form 
\begin{equation}
d_{\alpha\beta}(\omega_{s},\omega_{i}) = d(\omega_{s},\omega_{i})\left(\begin{array}{cc} 
 \hat{x}\hat{x} & 0\\
0 & \hat{y}\hat{y}
\end{array}\right)_{\alpha\beta}. 
\label{dI}
\end{equation}
Here, $\hat{x}$ and $\hat{y}$ are the unit vectors for the polarization state of the signal and idler photons and are associated with the tensor indices $\alpha$ and $\beta$. The contraction of the $TE$ and $TM$ vector wave functions with such a
susceptibility are given in Appendix~\ref{App:Dyad}. In this geometry the
cross-terms between $\bm{M}(\mathbf{k}_{\perp})$ and
$\bm{N}(\mathbf{k}_{\perp})$ cancel. 

Performing the $z$-integral over the length of the crystal and retaining only
those terms that are close to perfect phase matching one finds
\begin{align}
A_{\lambda\mu} &= \frac{\hbar E_{p}L}{4\pi i\varepsilon_0}
\frac{\omega_{s}^{2}\omega_{i}^{2}}{c^{4}}e^{iq_{p}z_{p}}
d(\omega_{s},\omega_{i})A^{\ast}(\omega_{s})A^{\ast}(\omega_{i})
I_{\lambda\mu}^{(I)}
\label{ampTI}
\end{align}
with
\begin{align}
I_{\lambda\mu}^{(I)} &= \int\frac{d^{2}k_{\perp}}{(2\pi)^2}
\frac{1}{k_{\perp}^{2}}\frac{1}{k_{z,s}k_{z,i}}
\left[\frac{\sin{\Delta k\frac{L}{2}}}{\Delta k\frac{L}{2}}\right]
e^{i\Sigma k\frac{L}{2}}\nonumber\\
&\times e^{i\mathbf{k}_{\perp}\cdot
(\mathbf{r}_{ds,\perp}-\mathbf{r}_{di,\perp})}
e^{iq_{z,s}z_{ds}}e^{iq_{z,i}z_{di}}\nonumber\\
&\times\bigg[M_{\lambda}(\mathbf{k}_{\perp})M_{\mu}(-\mathbf{k}_{\perp})
X_{TE,TE}(\omega_{s},\omega_{i})\nonumber\\
&+\left(\frac{k_{z,s}k_{z,i}}{k_{s}k_{i}}\right)
N_{\lambda}(\mathbf{k}_{\perp})N_{\mu}(-\mathbf{k}_{\perp})
X_{TM,TM}(\omega_{s},\omega_{i})\bigg].
\label{I}
\end{align}
Here,
\begin{align}
X_{\sigma\sigma'}(\omega_{s},\omega_{i}) &=
t^{12}_{TEM}(\omega_{p})t^{23}_{\sigma}(\omega_{s})
t^{23}_{\sigma'}(\omega_{i})\nonumber\\
&\times
M_{TEM}(\omega_{p})M_{\sigma}(\omega_{s})M_{\sigma'}(\omega_{i})\nonumber\\
&\times
\left(1+r^{23}_{TEM}(\omega_{p})r^{21}_{\sigma}(\omega_{s})
r^{21}_{\sigma'}(\omega_{i})e^{i\Sigma k L}\right),
\label{X}
\end{align}
with $\sigma,\sigma'\rightarrow TE,TM$ as required. $z_{ds}$ and $z_{di}$
are the perpendicular distances from the output face of the crystal to the
detectors. The longitudinal phase mismatch and longitudinal phase sum are 
\begin{align}
\Delta k = k_{p} - k_{z,s} -k_{z,i},\label{Dk}\\
\Sigma k = k_{p} + k_{z,s}+ k_{z,i},\label{Sk}
\end{align}
with the $z$-component and magnitude of the wave vector, respectively, being
$k_{z} = \sqrt{k^{2} - k^{2}_{\perp}}$, $k = n(\omega)\omega/c$ inside the
crystal and $q_{z} = \sqrt{q^{2} - k^{2}_{\perp}}$, $q=\omega^{2}/c^{2}$ in the
vacuum.

For a complete solution, the integral $I^{(I)}_{\lambda\mu}$ in Eq. \eqref{I}
would have to be solved numerically. However, in certain limits it is possible
to obtain an analytical solution. For example, let us consider the degenerate
($\omega_{s} = \omega_{i} = \omega_{p}/2 = \omega$, $k_{s} = k_{i} = k$),
collinear ($\mathbf{r}_{ds, \perp} = \mathbf{r}_{di, \perp}$) case. Expanding
the tensor product of the $TE$ and $TM$ vector wave functions as shown in
Appendix \ref{App:Dyad}, and performing the integral over the angular component
of $\mathbf{k}_{\perp}$, one finds that terms with odd powers of $k_{x}$ or
$k_{y}$ vanish. Hence,
\begin{align}
I_{\lambda\mu}^{(I)} &=
\int\frac{dk_{\perp}}{4\pi}\frac{k_{\perp}}{k_{z}^{2}}
\left[\frac{\sin{\Delta k\frac{L}{2}}}{\Delta k\frac{L}{2}}\right]
e^{i\Sigma k\frac{L}{2}}e^{iq_{z}(z_{ds}+z_{di})}
\mathbb{I}_{\lambda\mu}
\nonumber\\
&\times\bigg[X_{TE,TE}(\omega_{s},\omega_{i})+\left(\frac{k_{z}}{k}\right)^{4}
X_{TM,TM}(\omega_{s},\omega_{i})\bigg],
\end{align}
where $\mathbb{I}_{\lambda\mu}$ are the components of the 2-by-2 unit matrix.

An asymptotic approximation to the $k_\perp$-integral can be found 
in the far-field limit where one assumes that the distances from 
the crystal face to the detectors are large
compared to the wavelength of the detected modes, $z\omega/c \gg 1$. As, on
mere practical grounds, this is always the case, this limit provides a good
approximation for the count rate. However, one should note that for large
crystal-detector distances the propagation of the outgoing modes is effectively
parallel. Hence, in the far-field limit the count rate is asymptotic to the
collinear case. For non-collinear propagation one cannot use this approximation
and alternative methods for solving the integral must be found. 

In the far-field limit one can use the method of stationary phase to evaluate
the $k_{\perp}$-integral. At the stationary phase point at $k_{\perp}=0$ one
finds that $k_{z}=k=n(\omega)\omega/c$. Thus, the Fresnel coefficients become
\begin{gather}
r^{21/23}_{TE}(\omega) = -r^{21/23}_{TM}(\omega) = \frac{n(\omega)-1}{n(\omega)+1},\\
t^{23}_{TE}(\omega)  =  t^{23}_{TM}(\omega) =
\frac{2n(\omega)}{n(\omega)+1},\\ 
M_{TE}(\omega) = M_{TM}(\omega) = \left[1-r^{21}(\omega) r^{23}(\omega)
e^{2in(\omega)\frac{\omega}{c}L}\right]^{-1},
\end{gather}
and hence
\begin{align}
X_{TE,TE}(\omega,\omega) = X_{TM,TM}(\omega,\omega) = X_{+}(\omega,\omega),\label{X+}\\
X_{TE,TM}(\omega,\omega) = X_{TM,TE}(\omega,\omega) = X_{-}(\omega,\omega).\label{X-}
\end{align}
Furthermore,
\begin{align}
\Delta k &= \frac{2\omega}{c}[n(2\omega) - n(\omega)],\\
\Sigma k &= \frac{2\omega}{c}[n(2\omega) + n(\omega)].
\end{align}
The leading-order contribution to the integral gives
\begin{multline}
I_{\lambda\mu}^{(I)} = \frac{i}{2\pi}\frac{c}{\omega
n(\omega)^{2}}\frac{e^{i\frac{\omega}{c}(z_{ds}+z_{di})}}{(z_{ds}+z_{di})}\\
\times\left[\frac{\sin{\Delta k\frac{L}{2}}}{\Delta k\frac{L}{2}}\right]
e^{i\Sigma k\frac{L}{2}}
X_{+}(\omega,\omega) 
\mathbb{I}_{\lambda\mu}
\label{farTI}
\end{multline}
Substituting  Eq.~\eqref{farTI} back into Eq.~\eqref{ampTI}, one finds that
the biphoton amplitude for Type I conversion is given by
\begin{multline}
A_{\lambda\mu} =
\frac{\hbar E_{p}L}{8\pi^{2}\varepsilon_0}\frac{\omega^{3}}{c^{3}n(\omega)^{2}} 
d(\omega,\omega)
\frac{e^{i\frac{\omega}{c}(2z_{p}+z_{ds}+z_{di})}}{(z_{ds}+z_{di})}
\\ \times A^{\ast}(\omega)^{2}
\left[\frac{\sin{\Delta k\frac{L}{2}}}{\Delta k\frac{L}{2}}\right]
e^{i\Sigma k\frac{L}{2}}X_{+}(\omega,\omega)\mathbb{I}_{\lambda\mu},
\label{ampFinalI}
\end{multline}
with the corresponding coincident count rate
\begin{multline}
R^{(I)}(\omega) = \frac{\hbar^2
\omega^{6}L^{2}}{32\pi^{4}\varepsilon_0^{2}c^{6}|n(\omega)|^{4}}
\frac{|E_{p} d(\omega,\omega)|^2}{(z_{ds}+z_{di})^2}|A(\omega)|^{4}\\
\times |X_{+}(\omega,\omega)|^{2}\left|
\left[\frac{\sin{\Delta k\frac{L}{2}}}{\Delta k\frac{L}{2}}\right]
e^{i\Sigma k\frac{L}{2}}\right|^{2}.
\label{rateI}
\end{multline}

\subsection{Type II down conversion}
\label{Type II}

Similarly, for Type II down conversion the Green tensors in Eq.~\eqref{amp4} are
expanded in terms of vector wave functions as shown
in Appendix~\ref{App:Green} and the integral over the transverse
components of $\mathbf{r}_{A}$ is performed. In Type II conversion the
polarization unit vectors of the outgoing modes are perpendicular. Hence the
nonlinear susceptibility has the form 
\begin{equation}
d_{\alpha\beta}(\omega_{s},\omega_{i}) = d(\omega_{s},\omega_{i})\left(\begin{array}{cc} 
0 & \hat{x}\hat{y}\\
 \hat{y}\hat{x} & 0
\end{array}\right)_{\alpha\beta}. 
\label{dII}
\end{equation}
As before, $\hat{x}$ and $\hat{y}$ are the polarization unit vectors associated with the tensor indices $\alpha$ and $\beta$. The contraction of the $TE$ and $TM$ vector wave functions are again given in
Appendix~\ref{App:Dyad}. One should note that for this geometry the cross-terms
between the $\bm{M}(\mathbf{k}_{\perp})$ and $\bm{N}(\mathbf{k}_{\perp})$ $TE$
and $TM$ vector wave function do contribute. 

The $z$-integral is performed as before, keeping only those terms that are close
to perfect phase matching. Thus, the biphoton amplitude becomes
\begin{align}
A_{\lambda\mu} &= \frac{\hbar E_{p}}{4\pi i\varepsilon_0}
\frac{\omega_{s}^{2}\omega_{i}^{2}}{c^{4}}e^{iq_{p}z_{p}}
d(\omega_{s},\omega_{i})A^{\ast}(\omega_{s})A^{\ast}(\omega_{i})
I_{\lambda\mu}^{(II)}
\label{ampTII}
\end{align}
with
\begin{align}
&I_{\lambda\mu}^{(II)} = \int\frac{d^{2}k_{\perp}}{(2\pi)^2}
\frac{1}{k_{\perp}^{4}}\frac{1}{k_{z,s}k_{z,i }} 
\left[\frac{\sin{\Delta k\frac{L}{2}}}{\Delta k\frac{L}{2}}\right]
e^{i\Sigma k\frac{L}{2}}\nonumber\\
&\times e^{i\mathbf{k}_{\perp}\cdot
(\mathbf{r}_{ds,\perp}-\mathbf{r}_{di,\perp})}
e^{iq_{z,s}z_{ds}}e^{iq_{z,i}z_{di}} \nonumber\\
&\times\bigg[-2k_{x}k_{y}M_{\lambda}(\mathbf{k}_{\perp})
M_{\mu}(-\mathbf{k}_{\perp})X_{TE,TE}(\omega_{s},\omega_{i})\nonumber\\
&-i(k^{2}_{x}-k^{2}_{y})\left(\frac{k_{z,s}}{k_{s}}\right)
N_{\lambda}(\mathbf{k}_{\perp})M_{\mu}(-\mathbf{k}_{\perp})
X_{TM,TE}(\omega_{s},\omega_{i} )\nonumber\\
&+i(k^{2}_{x}-k^{2}_{y})\left(\frac{k_{z,i}}{k_{i}}\right)
M_{\lambda}(\mathbf{k}_{\perp})N_{\mu}(-\mathbf{k}_{\perp})
X_{TE,TM}(\omega_{s},\omega_{i} )\nonumber\\
&+2k_{x}k_{y}\left(\frac{k_{z,s}k_{z,i}}{k_{s}k_{i}}\right)
N_{\lambda}(\mathbf{k}_{\perp})N_{\mu}(-\mathbf{k}_{\perp})
X_{TM,TM}(\omega_{s},\omega_{i})\bigg],
\label{II}
\end{align}
with $X_{\sigma\sigma'}$, $\Delta k$ and $\Sigma k$ defined in Eq.~\eqref{X},
Eq.~\eqref{Dk} and Eq.~\eqref{Sk}.

For the degenerate, collinear case, after expanding the tensor product of
the $TE$ and $TM$ vector wave functions (Appendix~\ref{App:Dyad}), and
performing the angular integral, one finds
\begin{align}
I_{\lambda\mu}^{(II)} &=
\int\frac{dk_{\perp}k_{\perp}}{8\pi k_{z}^{2}}
\left[\frac{\sin{\Delta k\frac{L}{2}}}{\Delta k\frac{L}{2}}\right]
e^{i\Sigma k\frac{L}{2}}e^{iq_{z}(z_{ds}+z_{di})}
J_{\lambda\mu}
\nonumber\\
&\times\bigg[X_{TE,TE}(\omega,\omega)+\left(\frac{k_{z}}{k}\right)^{2}
X_{TM,TE}(\omega,\omega)\nonumber\\
&+\left(\frac{k_{z}}{k}\right)^{2}X_{TE,TM}(\omega,\omega)
+\left(\frac{k_{z}}{k}\right)^{4}X_{TM,TM}(\omega,\omega)\bigg],
\end{align}
with the leading-order contribution to the far-field limit given by
\begin{multline}
I_{\lambda\mu}^{(II)} = \frac{i}{4\pi}\frac{c}{\omega n(\omega)^{2}}
\frac{e^{i\frac{\omega}{c}(z_{ds}+z_{di})}}{(z_{ds}+z_{di})}
\left[ \frac{\sin{\Delta k\frac{L}{2}}}{\Delta k\frac{L}{2}}\right]
e^{i\Sigma k\frac{L}{2}}\\
\times \big[X_{+}(\omega,\omega) + X_{-}(\omega,\omega)\big]
J_{\lambda\mu}.
\label{farTII}
\end{multline}
Here, $J_{\lambda\mu}$ are the components of the 2-by-2 exchange matrix. 
Substituting Eq.~\eqref{farTII} back into 
Eq.~\eqref{ampTII}, one finds the
biphoton amplitude for Type II conversion is given by
\begin{multline}
A_{\lambda\mu} = \frac{\hbar
E_{p}L}{16\pi^{2}\varepsilon_0}\frac{\omega^{3}}{c^{3}n(\omega)^{2}}
d(\omega,\omega)
\frac{e^{i\frac{\omega}{c}(2z_{p}+z_{ds}+z_{di})}}{(z_{ds}+z_{di})}
J_{\lambda\mu}
\\
\times A^{\ast}(\omega)^{2}
\left[\frac{\sin{\Delta k\frac{L}{2}}}{\Delta k\frac{L}{2}}\right]
e^{i\Sigma k\frac{L}{2}}\big[X_{+}(\omega,\omega) +
X_{-}(\omega,\omega)\big],
\label{ampFinalII}
\end{multline}
with the corresponding coincidence count rate
\begin{multline}
R^{(II)}(\omega) = \frac{\hbar^2
\omega^{6}L^{2}}{128\pi^{4}\varepsilon_0^{2}c^{6}|n(\omega)|^{4}}
\frac{|E_{p} d(\omega,\omega)|^2}{(z_{ds}+z_{di})^2}|A(\omega)|^{4}\\
\times \left|\big[X_{+}(\omega,\omega) +
X_{-}(\omega,\omega)\big]\right|^{2}\left| 
\left[\frac{\sin{\Delta k\frac{L}{2}}}{\Delta k\frac{L}{2}}\right]
e^{i\Sigma k\frac{L}{2}}\right|^{2}.
\label{rateII}
\end{multline}

\section{Discussion}
\label{Dis}

\begin{figure}[ht]
\centering
\includegraphics[width=0.9\linewidth]{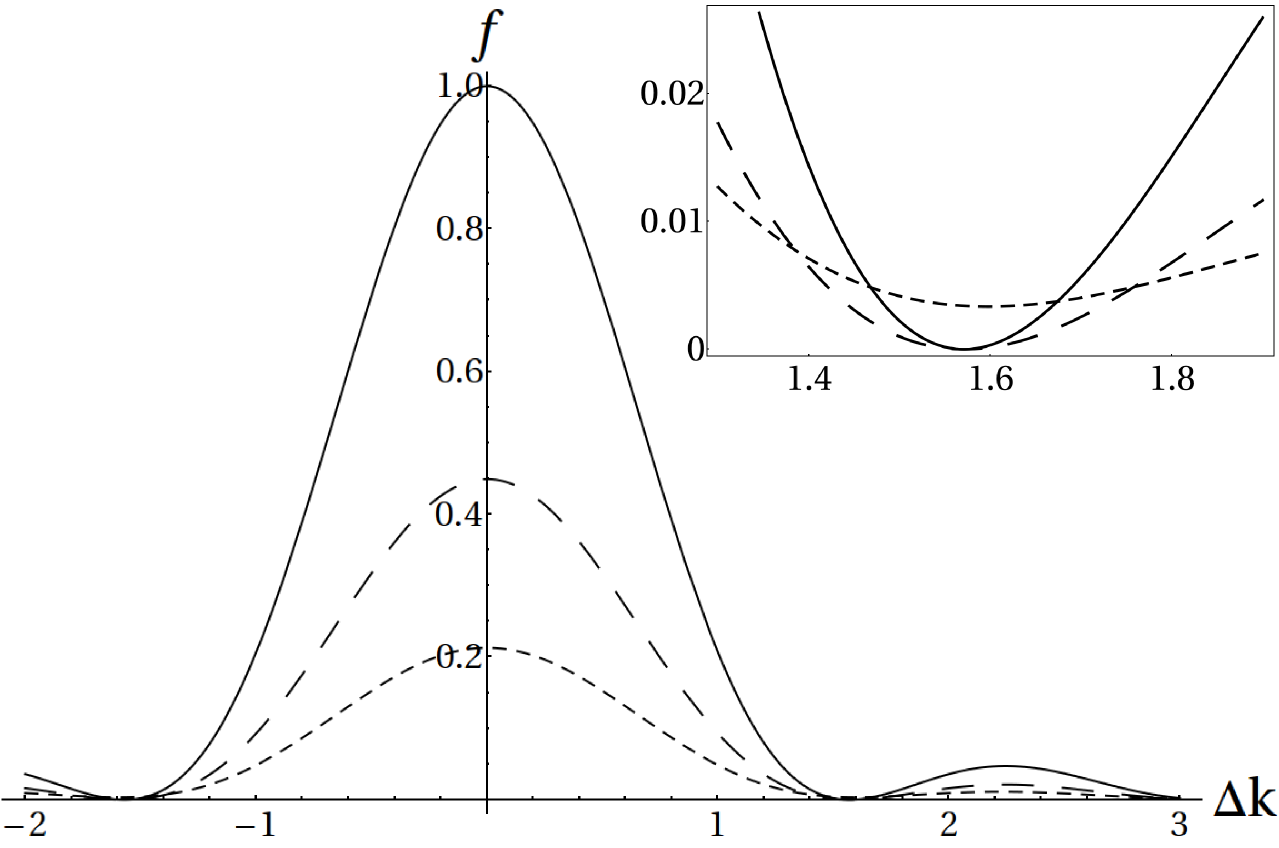}
\caption[Schematic plot of phase mismatched count rates for varying levels of
absorption.]{Dependence of the longitudinal phase matching condition
for vanishing absorption (solid line), non-vanishing absorption (small dashed
line) and the special case of degenerate, frequency independent absorption
(large dashed line).
$f=|\mathrm{sinc}\left[\Delta k\frac{L}{2}\right]e^{i\Sigma k\frac{L}{2}}|^{2}$.
The inset shows an expanded view of the first minimum. The units on both axes
are arbitrary.}
\label{Sinc} 
\end{figure}

Before discussing the details of the count rates in Eqs.~\eqref{rateI} and
\eqref{rateII}, there are a few general features worth commenting on. Both Type
I and Type II rates are proportional to the pump intensity, the square of the
non-linear susceptibility and have a `sinc'-function dependence on the
longitudinal phase matching condition, all of which are expected features of PDC
rates. Whilst the first two aspects are unaffected by absorption, the
`sinc'-function, in certain cases, can be distorted (see Fig.~\ref{Sinc}). 

In the presence of absorption, in addition to the expected reduction in
amplitude one finds a raising of the minimum. This is caused by the reduction in
intensity of one mode compared to the other, which leads to incomplete
destructive interference. In the special case of degenerate conversion and
identical absorption at all frequencies one sees complete destructive
interference because, in this case, the amplitude of all modes are reduced
equally. 

The inclusion of absorption adds two new features to the count rate.
Firstly, there are additional interaction terms, such as the nonlinear
interaction of the electric field with the noise polarization field. Secondly,
one observes the expected exponential decay of the modes via linear absorption. 
In the following we will consider the effects of PDC in beta barium borate, 
$\beta-BaB_{2}O_{4}$ (BBO), a common nonlinear crystal that can produce both 
Type I and Type II down conversion depending one the orientation of the pump 
beam to the crystal axis. In reality, BBO is strongly birefringent (c.f. Table 
\ref{BBO}, see also Ref.~\cite{bbo}). However, for our example we neglect this
aspect as it does not affect the features that are due to absorption; the
features we wish to highlight in this article. Thus, we take the refractive
index $n(\omega)$ to be equal to the values of the ordinary wave,
$n(\omega_{o})$, as opposed to its value of the extraordinary wave,
$n(\omega_{e})$.
\begin{table}[b]
\centering                    
\begin{tabular}{c c c} 
\hline                           
$\lambda (\mathrm{nm})$ & $Re[n(\omega_{o})]$ & $Re[n(\omega_{e})]$\\ 
\hline
$1064$ & $1.65$ & $1.54$ \\[1ex]  
$532$ & $1.67$ & $1.55$ \\[1ex]  
$266$ & $1.75$ & $1.61$ \\[1ex]  
\hline         
\end{tabular}
\caption {Table of frequency dependent refractive indices for BBO \cite{bbo}.} 
\label{BBO}
\end{table}

\begin{figure}[ht]
\centering
\includegraphics[width=0.9\linewidth]{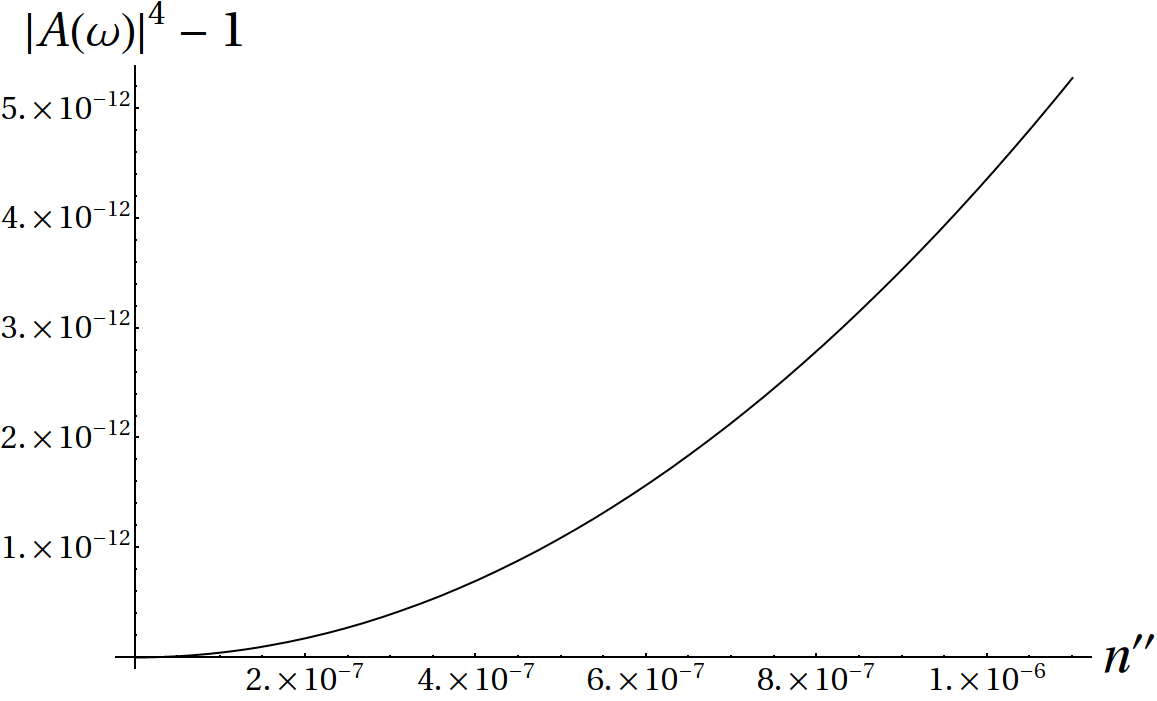}
\caption[Plot of relative count rate change owing to the nonlinear noise field
interactions.]{Relative increase of $A(\omega)$ with absorption. $n''$ is the
imaginary part of the refractive index. The real part of the refractive index is
taken to be $Re[n] = 1.67$.}
\label{AGraph} 
\end{figure}

Let us first consider the effect of the extra interaction terms, which refer to
the down-conversion of the pump field to one or more noise polarization fields.
From Eqs.~\eqref{amp2} and \eqref{amp3}, one observes that the contribution of
these extra interactions to the rate is contained in the $A(\omega)$ prefactors
as defined in Eq.~\eqref{A}. For vanishing absorption, $A(\omega)\rightarrow 1$.
As shown in Fig.~\ref{AGraph}, as absorption increases, so does $A(\omega)$.
The increase is due to the extra conversion pathways available to the pump
mode. This increase in final-state phase space causes an increase in the
probability of conversion and hence 
the rate. This kind of transition enhancement effect has already been seen in
linear systems, for example in the surface-induced transition rates in
Refs.~\cite{acta,decay}. For significant absorption of around $10\%$ per cm we
expect a relative increase in the count rate on the order of $5 \times 10^{-12}$.

\begin{figure}[!h]
\centering
\includegraphics[width=0.9\linewidth]{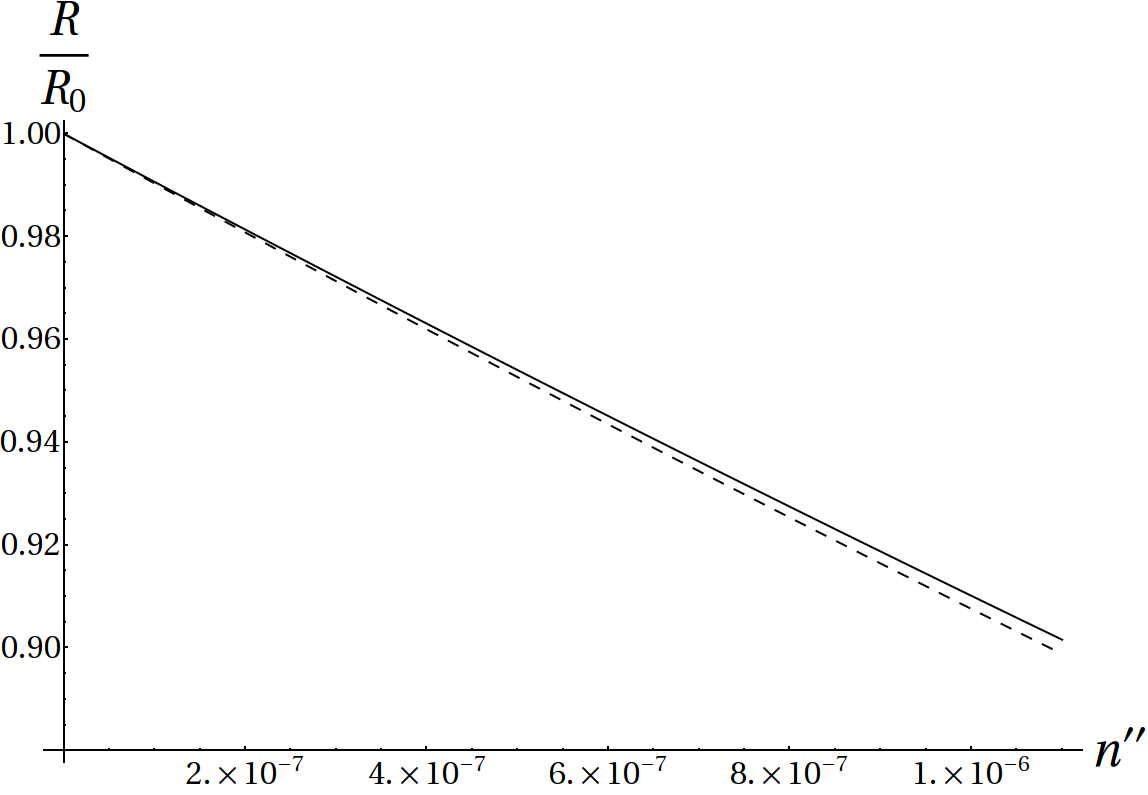}
\caption[Plot of the relative change of the coincident count rate with
absorption.]{Relative change of the coincident count rate with absorption for
Type I (solid line) and Type II (dashed line) conversion for a crystal
length of $L=2mm$. Here, $Re[n(2\omega)] = 1.75$, $Re[n(\omega)] = 1.67$,
$\omega = 3.54 \times 10^{15}s^{-1}$. The imaginary part of the refractive index
$n''$ is taken to be frequency independent. $R_{0}$ is the rate for vanishing
absorption.}
\label{Rates} 
\end{figure}

\begin{figure}[ht]
\centering
\subfigure[Type I Conversion]{
\includegraphics[width=0.9\linewidth]{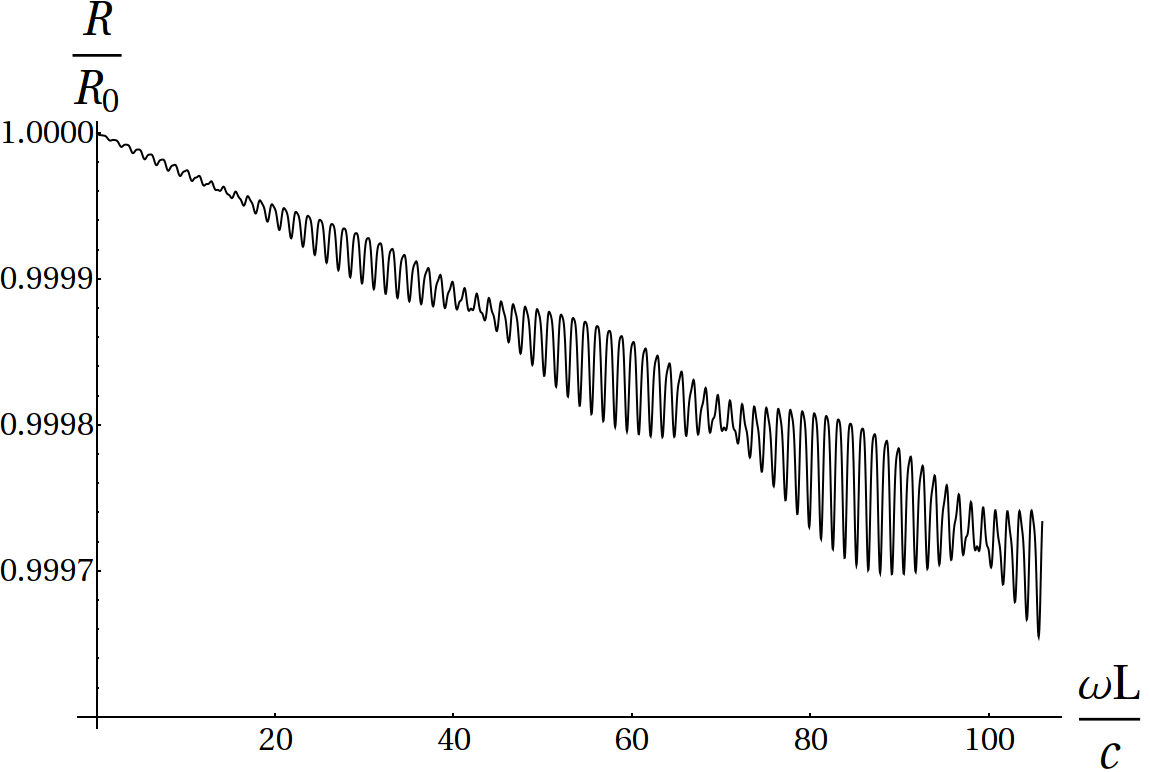}
}
\subfigure[Type II Conversion]{
\includegraphics[width=0.9\linewidth]{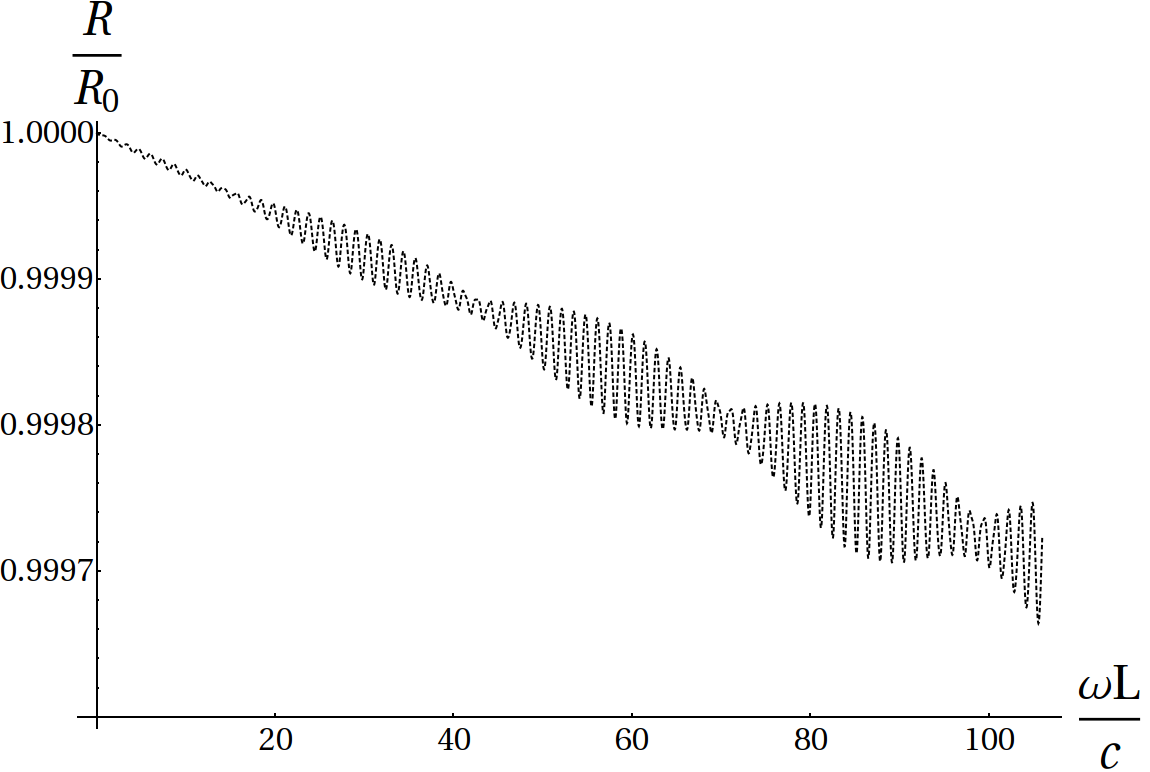}
}
\caption[Plot of interference between modes within a nonlinear crystal.]{The
interference of the modes within the crystal for Type I and Type II down
conversion. The beat frequencies occur with changing crystal length as the
reflected mode moves in and out of phase with the propagating mode.
Parameters as above.
}
\label{beats}
\end{figure}

Superimposed on the small rate increase caused by the extra interaction terms
is the effect of linear absorption on the propagating modes. Of the two effects,
this is by far the more dominant. Figure~\ref{Rates} shows the relative change
in the count rate as absorption is increased. 
Notice that the Type I and Type II rates differ. This difference is due to the
interference between the pump beam and the outgoing modes, and between forward
propagating modes and those which have been reflected from the crystal
interfaces [c.f. Eqs.~\eqref{X+}, \eqref{X-}, \eqref{rateI}, and
\eqref{rateII}]. In fact, as one changes the crystal length the relative rates
of Type I and Type II change, with some lengths resulting in a larger Type II
rate compared to Type I. One also observes the appearance of beat frequencies in
the count rate as the crystal length is changed (see Fig.~\ref{beats}). These
are also caused by interference of the modes within the crystal. Note also that
a longer crystal length leads to a larger distance over which linear absorption
can act, hence the general reduction in the count rate with increasing crystal
length.

Let us comment on a few observations. Firstly, owing to the dyadic structure of the Green tensor,
the detected modes and the modes created at the interaction point decouple.
Thus, the count rate of a specific polarization state is a function of both
possible polarization states [c.f. Eqs.~\eqref{dI} and \eqref{dII}]. This
highlights a fundamental principle of quantum mechanics that, until measured,
the signal and idler modes propagate as a superposition of both possible
polarization states. Another observation is that the contributions to the count
rate from each of the detected polarization states are equal [c.f.
Eqs.~\eqref{ampFinalI} and \eqref{ampFinalII}]. It is easy
to show that, despite absorption, this is a maximally entangled
state with respect to polarization. In fact, since the outgoing modes propagate
as both polarization states, even with a more complicated material, (e.g
birefringence, frequency dependent absorption, etc.) the detected state will
still be maximally entangled. Similar arguments can be used show that the
Hong-Ou-Mandel interference pattern from two-photon interference is unaffected
and it is still possible to get maximal visibility despite absorption. It is
important to note that this analysis has described a experimental setup that
post-selects the two photon state. It is possible that during the propagation
through the crystal one photon of the down-converted pair is absorbed and lost
completely. However, these states are not detected in coincident counting
experiments and hence do not affect the results outlined above.

\section{Summary}
\label{Sum}

In this article we have investigated the coincident count rates for Type I
and Type II parametric down-conversion, subject to absorption. Despite initial 
assumptions to the contrary, we found that the increase in rate that is caused 
by the extra nonlinear interactions of the electric field with the noise 
polarization field are negligible compared to the effect of linear absorption. 
Thus we have conclusively shown that nonlinear absorption and nonlinear noise 
interactions can be safely neglected in all current experimental work. We also 
found that the rates for Type I and Type II conversion differ. Aside from the 
possibility of differing nonlinear susceptibilities the two types of conversion, 
they also differ owing to the different interference processes that occur between 
the waves that scattered from the crystal boundaries. 

We also observe that, in spite of absorption, the entanglement and two-photon
interference properties of the photon pairs are unchanged by both the extra
nonlinear noise interactions and by linear absorption if one assumes that both
photons are detected. However, absorption will play an important role in single
photon counting experiments where one only detects one of the photons in the
pair. Unlike in coincident counting experiments, where the loss of one photon
means that the photon pair is undetected, in the case of single photon counting
these partially absorbed photon pairs will contribute significantly to the count
rate. Hence, in single photon counting experiments both linear absorption and
nonlinear noise interactions will play a more important role. 

\acknowledgments

This work was supported by the UK Engineering and Physical Sciences
Research Council. One of us (JAC) would like to thank C. Kurtsiefer and M. Tame for 
useful insight into various aspects of the PDC process.

\appendix

\section{The Green tensor for planar multilayered media}
\label{App:Green}

We consider a dielectric medium that is layered in the $z$-direction and
consists of three regions (vacuum-crystal-vacuum). Each region is infinitely
extended in the $(x,y)$-plane. The interfaces between the crystal and the vacuum
are located at $z=\pm L/2$. The scattering Green tensor that describes
transmission through the output face of a crystal at $z=+L/2$ is given by
\cite{chew}
\begin{align}
G_{\alpha\beta}(\mathbf{r}_{d},\mathbf{r}_{A},\omega) &=
\frac{i}{2(2\pi)^2}\int
\frac{d^{2}k_{\perp}}{k_{\perp}^{2}}\frac{1}{k_{z}}e^{i\mathbf{k}_{\perp}
\cdot(\mathbf{r}_{d, \perp}-\mathbf{r}_{A, \perp})}\nonumber\\
&\times\big[M_{\alpha}(\mathbf{k}_{\perp})
M_{\beta}(-\mathbf{k}_{\perp})F^{23}_{TE}(z_{d},z_{A})\nonumber\\
&+N_{\alpha}(\mathbf{k}_{\perp})N_{\beta}(-\mathbf{k}_{\perp})
F^{23}_{TM}(z_{d},z_{A})\big],
\end{align}
with $TE$ and $TM$ vector wave functions
\begin{equation}
\bm{M}(\mathbf{k}_{\perp}) = i(\mathbf{k}_{\perp}\times\hat{z}),\quad
\bm{N}(\mathbf{k}_{\perp}) =
-\frac{1}{k}(\mathbf{k}_{j\pm}\times\mathbf{k}_{\perp}\times\hat{z}).
\end{equation}
Here $\mathbf{r}_{\perp}=(x,y,0)$ and $\mathbf{k}_{\perp}=(k_{x},k_{y},0)$
are vectors restricted to the $(x,y)$-plane, $k_{\perp} = |\mathbf{k}_{\perp}|$
and $k_{z}=\sqrt{k^{2}-k^{2}_{\perp}}$ with
$k=n(\mathbf{r}_{A},\omega)\omega/c$ being the magnitude of the wave vector
in the crystal. The vector $\mathbf{k}_{j\pm}$ is defined as
$\mathbf{k}_{j\pm}=\mathbf{k}_{\perp}\pm k_{z}\hat{z}$ where the upper (lower)
sign is taken for $z_{d} > z_{A}$ ($z_{d}<z_{A}$). As $z_{A}$ and $z_d$ are
assumed to be in region $2$ and region $3$, respectively, one always has
$z_{d}>z_{A}$. Hence, in this case the positive sign is valid. 

The $z$-dependent factors are given by
\begin{multline}
F^{(23)}_{\sigma}(z_{d},z_{A}) =
t^{23}_{\sigma}(\omega)e^{iq_{z}(z_{d}-\frac{L}{2})}e^{ik_{z}\frac{L}{2}}\\
\times\big[e^{-ik_{z}z_{A}} + r^{21}_{\sigma}(\omega)e^{ik_{z}(z_{A}+L)}\big]
M_{\sigma}(\omega).
\label{F}
\end{multline}
with $\sigma \rightarrow TE, TM$. The relevant Fresnel coefficients are
\begin{gather}
r^{21/23}_{TE}(\omega) = \frac{k_{z}-q_{z}}{k_{z}+q_{z}}, \quad
r^{21/23}_{TM}(\omega) =
\frac{k_{z}-\varepsilon(\omega)q_{z}}{k_{z}+\varepsilon(\omega)q_{z}},
\nonumber\\
t^{23}_{TE}(\omega)  = \frac{2k_{z}}{k_{z}+q_{z}}, \quad t^{23}_{TM}(\omega) =
\frac{2k_{z}}{k_{z}+\varepsilon(\omega)q_{z}}.
\end{gather}
with the magnitude of the wave vector in vacuum 
$q_{z} =\sqrt{q^{2}-k^{2}_{\perp}}$ with $q=\omega/c$. The function
\begin{equation}
M_{TE/TM}(\omega) =
\big[1-r^{21}_{TE/TM}(\omega)r^{23}_{TE/TM}(\omega)e^{2ik_{z}L}\big]^{-1}
\end{equation}
accounts for multiple reflections inside the crystal.

\section{Commutation relations}
\label{App:Comm}

The commutation relations for the macroscopic fields can be found from the
bosonic expansions of the electric and noise polarization fields, Eq.~\eqref{E}
and Eq.~\eqref{P}, respectively. On application of Eq.~\eqref{comm} the
commutators for the macroscopic fields are found to be
\begin{gather}
\left[\hat{E}_{\alpha}(\mathbf{r},\omega),
\hat{E}_{\beta}^{\dagger}(\mathbf{r}',\omega')\right] =
\frac{\hbar\omega^{2}}{\pi\varepsilon_{0}c^{2}}
\mathrm{Im}G_{\alpha\beta}(\mathbf{r},\mathbf{r}',\omega)\delta_{\omega\omega'},
\\
\left[\hat{E}_{\alpha}(\mathbf{r},\omega),
\hat{P}_{N,\beta}^{\dagger}(\mathbf{r}',\omega')\right] =
\frac{\hbar\omega^{2}}{\pi c^{2}}\varepsilon''(\mathbf{r}',\omega)
G_{\alpha\beta} (\mathbf{r},\mathbf{r}',
\omega)\delta_{\omega\omega'},\\
\left[\hat{P}_{N,\alpha}(\mathbf{r},\omega),
\hat{E}^{\dagger}_{\beta}(\mathbf{r}',\omega')\right] =
\frac{\hbar\omega^{2}}{\pi c^{2}}\varepsilon''(\mathbf{r},\omega)
G^{\ast}_{\alpha\beta}(\mathbf{r},\mathbf{ r}',\omega)\delta_{\omega\omega'},\\
\left[\hat{P}_{N, \alpha}(\mathbf{r},\omega),
\hat{P}^{\dagger}_{N,\beta}(\mathbf{r}',\omega')\right] =
\frac{\hbar}{\pi}\varepsilon_{0}\varepsilon''(\mathbf{r},\omega)
\delta(\mathbf{r} - \mathbf{r}')\delta_{\alpha\beta}\delta_{\omega\omega'}.
\end{gather}

\section{Products of vector wave functions}
\label{App:Dyad}

The contraction of the $TE$ and $TM$ vector wave functions with the
susceptibility for a Type I process, $d_{\alpha\beta}(\omega_{s},\omega_{i}) =
d(\omega_{s},\omega_{i})\big[\hat{x}\hat{x}+\hat{y}\hat{y}\big]_{\alpha\beta}$,
leads to the following expressions
\begin{gather}
M_{s,\alpha}(-\mathbf{k}_{\perp})\big[\hat{x}\hat{x}+\hat{y}\hat{y}\big]_{\alpha\beta}M_{i,\beta}(\mathbf{k}_{\perp}) = k_{\perp}^{2},\nonumber\\ \\
M_{s,\alpha}(-\mathbf{k}_{\perp})
\big[\hat{x}\hat{x}+\hat{y}\hat{y}\big]_{\alpha\beta}
N_{i,\beta}(\mathbf{k}_{\perp}) = 0,\nonumber\\ \\
N_{s,\alpha}(-\mathbf{k}_{\perp})
\big[\hat{x}\hat{x}+\hat{y}\hat{y}\big]_{\alpha\beta}
M_{i,\beta}(\mathbf{k}_{\perp}) = 0,\nonumber\\ \\
N_{s,\alpha}(-\mathbf{k}_{\perp})
\big[\hat{x}\hat{x}+\hat{y}\hat{y}\big]_{\alpha\beta}
N_{i,\beta}(\mathbf{k}_{\perp}) =
k_{\perp}^{2}\left(\frac{k_{z,s}k_{z,i}}{k_{s}k_{i}}\right).
\end{gather}
Similarly, for the susceptibility for a Type II process,
$d_{\alpha\beta}(\omega_{s},\omega_{i}) =
d(\omega_{s},\omega_{i})\big[\hat{x}\hat{y}+\hat{y}\hat{x}\big]_{\alpha\beta}$,
one finds
\begin{gather}
M_{s,\alpha}(-\mathbf{k}_{\perp})
\big[\hat{x}\hat{y}+\hat{y}\hat{x}\big]_{\alpha\beta}
M_{i,\beta}(\mathbf{k}_{\perp}) = -2k_{x}k_{y},\nonumber\\ \\
M_{s,\alpha}(-\mathbf{k}_{\perp})
\big[\hat{x}\hat{y}+\hat{y}\hat{x}\big]_{\alpha\beta}
N_{i,\beta}(\mathbf{k}_{\perp}) =
i(k_{x}^2-k_{y}^2)\left(\frac{k_{z,i}}{k_{i}}\right),\\
N_{s,\alpha}(-\mathbf{k}_{\perp})
\big[\hat{x}\hat{y}+\hat{y}\hat{x}\big]_{\alpha\beta}
M_{i,\beta}(\mathbf{k}_{\perp}) =
-i(k_{x}^2-k_{y}^2)\left(\frac{k_{z,s}}{k_{s}}\right),\\
N_{s,\alpha}(-\mathbf{k}_{\perp})
\big[\hat{x}\hat{y}+\hat{y}\hat{x}\big]_{\alpha\beta}
N_{i,\beta}(\mathbf{k}_{\perp}) =
2k_{x}k_{y}\left(\frac{k_{z,s}k_{z,i}}{k_{s}k_{i}}\right).
\end{gather}
The (dyadic) tensor products of the uncontracted vector wave functions are
\begin{align}
M_{s,\alpha}(\mathbf{k}_{\perp})&M_{i,\beta}(-\mathbf{k}_{\perp}) =
\left(\begin{array}{ccc} k_{y}^2 & -k_{x}k_{y} & 0 \\ -k_{x}k_{y} & k_{x}^2 & 0
\\ 0 & 0 & 0 \end{array}\right),\nonumber\\ \\
M_{s,\alpha}(\mathbf{k}_{\perp})&N_{i,\beta}(-\mathbf{k}_{\perp}) =\nonumber\\
-\frac{1}{k_{i}}&\left(\begin{array}{ccc} 
ik_{x}k_{y}k_{z,i} & ik_{y}^{2}k_{z,i} & -ik_{y}k^{2}_{\perp} \\ 
-ik_{x}^{2}k_{z,i} & -ik_{x}k_{y}k_{z,i} & ik_{x}k^{2}_{\perp} \\ 
0 & 0 & 0 \end{array}\right),\nonumber\\ \\
N_{s,\alpha}(\mathbf{k}_{\perp})&M_{i,\beta}(-\mathbf{k}_{\perp}) =\nonumber\\
\frac{1}{k_{s}}
&\left(\begin{array}{ccc} 
ik_{x}k_{y}k_{z,s} & -ik_{x}^{2}k_{z,i} & 0 \\ 
ik_{y}^{2}k_{z,s} & -ik_{x}k_{y}k_{z,i} & 0 \\ 
-ik_{y}k_{\perp}^2 & ik_{x}k_{\perp}^2 & 0
\end{array}\right),\nonumber\\ \\
N_{s,\alpha}(\mathbf{k}_{\perp})&N_{i,\beta}(-\mathbf{k}_{\perp}) =\nonumber\\-\frac{1}{k_{s}k_{i}}&\left(\begin{array}{ccc} 
k^{2}_{x}k_{z,s}k_{z,i} & k_{x}k_{y}k_{z,s}k_{z,i} & -k_{x}k_{z,s}k^{2}_{\perp} \\ 
k_{x}k_{y}k_{z,s}k_{z,i} & k^{2}_{y}k_{z,s}k_{z,i} & -k_{y}k_{z,s}k^{2}_{\perp} \\ 
-k_{x}k_{z,i}k^{2}_{\perp} & -k_{y}k_{z,i}k^{2}_{\perp} & k^{4}_{\perp}
\end{array}\right).\nonumber\\
\end{align}

\end{document}